\documentclass[twocolumn,preprintnumbers,amsmath,amssymb]{revtex4}
\usepackage{graphicx}
\usepackage{dcolumn}
\usepackage{bm}

\begin{document}
\title{
Feasibility of terahertz lasing in optically pumped epitaxial 
multiple  graphene layer structures 
}
\author{V.~Ryzhii,  M.~Ryzhii, and A. Satou
}
\affiliation{
Computational Nanoelectronics Laboratory, University of Aizu, 
Aizu-Wakamatsu 965-8580, Japan\\
Japan Science and Technology Agency, CREST, Tokyo 107-0075, Japan
}
\author{T. Otsuji}
\affiliation{
Research Institute of Electrical Communication, Tohoku University, Sendai 980-8577 and\\
Japan Science and Technology Agency, CREST, Tokyo 107-0075, Japan
}
\author{A. A. Dubinov and V. Ya. Aleshkin
}
\affiliation{
Institute for Physics of Microstructures, Russian Academy of Sciences, 
Nizhny Novgorod 603950, Russia
}
\date{\today}    
\begin{abstract}
A multiple-graphene-layer (MGL) structure with a stack of GLs and a highly 
conducting
bottom GL 
on SiC substrate pumped by optical radiation 
is considered as an active region of terahertz (THz) and far infrared (FIR)
lasers with external metal mirrors. The dynamic conductivity of the MGL structure 
is calculated as a function
of the signal frequency, the number of  GLs, and the optical 
pumping intensity. 
The utilization of optically pumped MGL structures might provide
the achievement of lasing  with the frequencies of about 1 THz at room
temperature 
due to a high efficiency.
\end{abstract}

\maketitle

\section{Introduction}

Since the first experimental demonstrations of the nontrivial properties of graphene
(see, for instance, Refs.~\cite{1,2} and numerous subsequent publications), 
due to the gapless energy spectrum of graphene, the latter can be used as an 
active region in the terahertz (THz) and far infrared (FIR) lasers with optical 
or injection pumping~\cite{3,4}. The optical pumping with the photon energy
$\hbar\Omega$ leads to the generation of electrons and holes with the energy
$\varepsilon_0 = \hbar\Omega/2$. Since the interaction of electrons and holes
with optical phonons is characterized by fairly short time $\tau_0$,
the photogenerated electrons and holes quickly  
emit   cascades of $N$ optical phonons, where $N = [\varepsilon_0/\hbar\omega_0]$
and $[X]$ means the integer part of $X$. 
Due to relatively high energy of optical phonons in graphene 
($\hbar\omega_0 \simeq 200$~meV), the number $N$ can vary from zero (pumping by 
CO$_2$ lasers) to a few units (pumping by quantum cascade lasers or semiconductor
injection diode lasers).
Thus, the photogenerated electrons and holes
populate the low energy regions of the conduction and valence bands 
of graphene~\cite{3} (see also Refs.~\cite{5,6}). 
As a result, the Fermi levels 
of electrons and holes  are separated and shifted from the Dirac 
point to the conduction and valence bands, respectively.
 This corresponds to  the population inversion for 
the interband transitions with
absorption or emission of photons with relatively low energies: 
$\hbar\omega < 2\varepsilon_{F}$, where $\varepsilon_F$ is the quasi-Fermi 
energy
of the electron and hole distributions, and leads to the 
negative contribution of the interband transitions to the real part of
the dynamic conductivity
Re~$\sigma_{\omega}$ (which includes the contributions of both interband and intraband transitions). 
If Re~$\sigma_{\omega} < 0$,
the stimulated emission of photons with the relatively low
energy $\hbar\omega$ (in the THz or FIR range) is possible.
The intraband contribution is primarily associated with the Drude mechanism
of the photon absorption. As shown previously, the realization
of the condition of lasing is feasible~\cite{3,6}. Some possible schemes
of the graphene lasers (with metal waveguide structure or external metal mirrors)
utilizing the above mechanism of optical pumping were
considered recently (see, for instance, Ref.~\cite{7}).
\begin{figure}[t]
\vspace*{-0.4cm}
\begin{center}
\includegraphics[width=6.5cm]{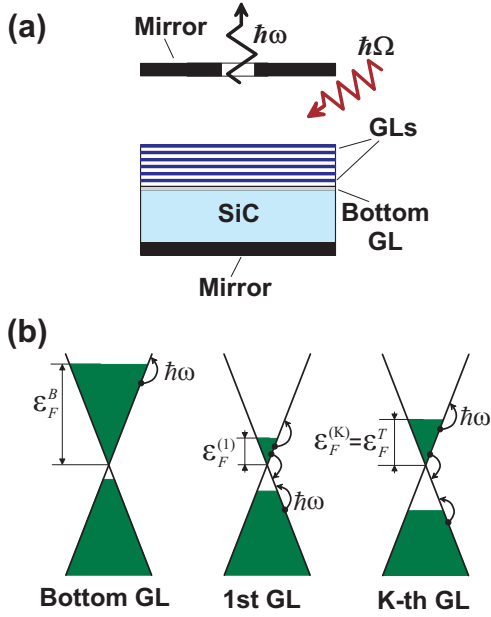}
\caption{(a) Schematic view of a laser with a MGL structure.
(b) Occupied (by electrons) and vacant states in different GLs under optical pumping.
Arrows show transitions related to interband emission and intraband
 absorption of THz photons with energy $\hbar\omega$ (interband transitions related to 
optical pumping as well the processes of intraband relaxation of 
the photogenerated electrons and holes  are not shown).
}
\end{center}
\end{figure}   

In particular,
the laser structure evaluated in Ref.~\cite{7}, includes two graphene
layers. Relatively weak absorption of optical radiation with the efficiency
$\beta = \pi e^2/\hbar\,c \simeq 0.023$, where $e$ is the electron charge,
$\hbar$ is the reduced Planck constant, and $c$ is the speed of light,
 per one passage necessitates  the use of rather strong
optical pumping. This drawback can be eliminated  in the structures 
with multiple
graphene layers (MGLs). Such epitaxial 
MGL structures including up to 100 very perfect
graphene layers with a high electron mobility
($\mu \simeq 250,000$~cm$^2$/sV) preserving up to the room
temperatures and, hence, with a long momentum relaxation time were recently 
fabricated 
using the thermal 
decomposition from a 4H-SiC substrate~\cite{8,9}.
The incident optical radiation can be almost totally absorbed in these MGL
 structures providing enhanced pumping efficiency.
A long momentum relaxation time (up to 20 ps~\cite{10}) implies 
that the intraband
(Drude) absorption in the THz and FIR ranges can be weak.
The optically pumped
MGL structures in question with the momentum relaxation time about several ps
at the room temperatures might be ideal active media for interband THz lasers.   
The situation, however, is complicated by the presence of highly 
conducting bottom GL
(and, hence, absorbing THz and FIR radiation due to the intraband processes)
near the the interface with SiC as a result of charge transfer from SiC.

In this paper, we analyze
the operation of  THz  lasers utilizing
optically pumped  MGL structures by calculating  their characteristics
as function of the number of graphene layers $K$ and optical pumping intensity
$I_{\Omega}$ and demonstrate the feasibility of realization of such lasers 
operating at the  frequencies from about 1 THz to several THz at room temperatures.

\section{Model}

We consider a laser structure with an MGL structure 
on a SiC substrate
serving
as its active region. Although the active media under consideration
can be supplemented by different resonant cavities, for  definiteness
we address to
the MGL structure  placed between the highly reflecting
metal mirrors (made of Al, Au, or Ag) as shown in Fig.~1(a).
It is assumed that the MGL structure under consideration comprises 
$K$ upper GLs (to which we refer to  just as GLs) and 
a highly 
conducting bottom layer
with a  Fermi energy of electrons $\varepsilon_F^B$, which  
is rather large:
$\varepsilon_F^B   \simeq 400$~meV ~\cite{8}. 
Apart from this, we briefly compare the MGL structure in question with that
in which the bottom GL is absent. The latter structure can be fabricated
by an additional peeling of upper GLs and depositing them on a Si substrate.

Due to the phogeneration
of electrons and holes with the energy $\varepsilon_0 = \hbar\Omega/2$
followed by their cooling associated with
 the cascade emission of optical phonons,
low energy
states near the bottom of the conduction band and 
the top of the valence band can be essentially occupied.
Taking into account that elevated
 electron and hole densities (at elevated temperatures and sufficiently strong optical pumping considered in the following), 
the electron and hole distributions  in  the range of energies
$\varepsilon \ll \hbar\omega_0$ in the $k-$th graphene layer 
($1 \leq k \leq K$) can be  described by
the Fermi  functions 
with the  
 quasi-Fermi energies
  $\varepsilon_F^{(k)}$ [see Fig.~1(b)].
The case of relatively weak pair collisions of the photogenerated
electrons and holes (due to their low densities)
in which the energy  distributions deviate  from the Fermi distributions
was studied in Ref.~\cite{6}. 
Using the Falkovsky-Varlamov formula~\cite{11} for the dynamic conductivity
of an MGL structure generalized for 
nonequilibrium electron-hole
systems~\cite{3}, one can obtain
$$
{\rm Re} \sigma_{\omega}^{B} =  \displaystyle
\biggl(\frac{e^2}{4\hbar}\biggr)\biggl\{1 - 
\biggl[1 + \exp\biggl(\frac{\hbar\omega/2 - \varepsilon_F^{B}}{k_BT}\biggr) 
\biggr]^{-1}
$$
$$
-\biggl[1 + \exp\biggl(\frac{\hbar\omega/2 + \varepsilon_F^{B}}{k_BT}\biggr) 
\biggr]^{-1}\biggr\}
$$
\begin{equation}\label{eq1}
+ 
\displaystyle\biggl(\frac{e^2}{4\hbar}\biggr)
\frac{4k_BT\tau_B}
{\pi\hbar(1 + \omega^2\tau_B^2)}
\ln \biggl[1 + \exp\biggl(\frac{\varepsilon_F^{B}}{k_BT}\biggr)\biggr]
\end{equation}
for the bottom GL,
and

$$
{\rm Re} \sigma_{\omega}^{(k)} =  \displaystyle
\biggl(\frac{e^2}{4\hbar}\biggr)
\tanh\biggl(\frac{\hbar\omega - 2\varepsilon_F^{(k)}}{4k_BT}\biggr)
$$
\begin{equation}\label{eq2}
+ \displaystyle\biggl(\frac{e^2}{4\hbar}\biggr)
\frac{8k_BT\tau}
{\pi\hbar(1 + \omega^2
\tau^2)}
\ln \biggl[1 + \exp\biggl(\frac{\varepsilon_F^{(k)}}{k_BT}\biggr)\biggr]
\end{equation}
for the GLs with $1 \leq k \leq K$. 
Here  $\tau_B$ and  $\tau$ are the electron and hole momentum relaxation 
times in the bottom and other GLs, respectively,
$T$ is the electron
and hole temperature, and $k_B$ is the Boltzmann constant. 
The first and second terms
in the right-hand sides of Eqs.~(1) and (2) correspond to the interband and 
intraband transitions, respectively, which are schematically shown
by arrows in Fig.~1(b). For simplicity we shall disregard
the variation of the electron and hole densities in the bottom GL under the 
optical pumping, so that the electron-hole system in this GL is assumed
to be close to equilibrium with the Fermi energy $\varepsilon_F^{B}$
determined by the interaction with the SiC substrate.

The quasi-Fermi energies in the GLs with $k \geq 1$ are mainly determined
by the electron (hole) density in  this layer $\Sigma_k$, i.e,
 $\varepsilon_F^{(k)}  \propto \sqrt{\Sigma^{(k)}}$ and, therefore,
by the rate of photogeneration  $G_{\Omega}^{(k)}$ 
by the optical
radiation (incident and reflected from the mirror) at the $k-$th GL plane.
Using Eq.~(2) for $\hbar\omega = \hbar\Omega$, we obtain
\begin{equation}\label{eq3}
G_{\Omega}^{(k)} =  
\frac{I_{\Omega}^{(k)}}{\hbar\Omega} 
\biggl(\frac{\pi\,e^2}{\hbar\,c}\biggr)
\tanh\biggl(\frac{\hbar\Omega - 2\varepsilon_F^{(k)}}{4k_BT}\biggr).
\end{equation}
Here $I_{\Omega}^{(k)}$ is the intensity (power density) of the optical pumping
radiation at the $k-$th GL. At $\hbar\Omega > 2\varepsilon_F^{(k)}$ (for all GLs),
Eq.~(3)
yields 
$G_{\Omega}^{(k)} \simeq 
\beta I_{\Omega}^{(k)}/\hbar\Omega$.
Considering the attenuation of the optical pumping radiation due to its absorption
in each GL, one can obtain

\begin{equation}\label{eq4}
G_{\Omega}^{(k)} =  
\frac{I_{\Omega}}{\hbar\Omega} \beta[(1 - \beta)^{K - k} + (1 - \beta_B)^2
(1 - \beta)^{K + k- 1} ].
\end{equation}
%
Here $I_{\Omega}$ is the intensity of incident pumping radiation
and $\beta_B = (4\pi/c) {\rm }\sigma_{\Omega}^B$.
The latter quantity accounts for the absorption of optical pumping radiation
in the bottom layer. 

A relationship between $\varepsilon_F^{(k)}$ and $G_{\Omega}^{(k)}$
is determined by the recombination mechanisms.
We assume that $\varepsilon_F^{(k)} \propto [G_{\Omega}^{(k)}]^{\gamma}$, where $\gamma$ is a phenomenological parameter.
In this case,
\begin{equation}\label{eq5}
\varepsilon_F^{(k)}= \varepsilon_F^{B}
\biggl[(1 - \beta)^{K-k}\frac{1 +  (1 - \beta_B)^2(1 - \beta)^{2k-1}}
{1 + (1 - \beta_B)^2(1 - \beta)^{2K-1}} 
\biggr]^{\gamma},
\end{equation}
where $\varepsilon_F^{B} = \varepsilon_F^{(K)}$ is the quasi-Fermi energy
in the topmost~GL.

\section{MGL structure net dynamic conductivity}

Taking into account that the thickness of the MGL structure
is small in comparison with the wavelength of THz/FIR radiation,
the generation and absorption of the latter is determined
by the real part of the net dynamic conductivity: 

\begin{equation}\label{eq6}
{\rm Re}~\sigma_{\omega}
= {\rm Re}~\sigma_{\omega}^{B} + {\rm Re}\sum_{k=1}^K\sigma_{\omega}^{(k)}.
\end{equation}
Taking into account that in the frequency range under consideration
$\hbar\omega \ll \varepsilon_F^B$, so that one can neglect
the first term in the right-hand side of Eq.~(1) responsible for the 
interband absorption in the bottom GL, 
 and 
using Eqs.~(1), (2), and (6), we arrive at
$$
{\rm Re}~\sigma_{\omega} = 
\displaystyle
\biggl(\frac{e^2}{4\hbar}\biggr)\biggl\{
\displaystyle
\frac{4k_BT\tau_B}
{\pi\hbar(1 + \omega^2\tau_B^2)}
\ln \biggl[1 + \exp\biggl(\frac{\varepsilon_F^{B}}{k_BT}\biggr)\biggr]
$$
$$
+ 
\displaystyle
\frac{8k_BT\tau}
{\pi\hbar(1 + \omega^2
\tau^2)}
\sum_{k = 1}^K\ln \biggl[1 + \exp\biggl(\frac{\varepsilon_F^{(k)}}{k_BT}\biggr)\biggr]
$$
\begin{equation}\label{eq7}
+
 \displaystyle
\sum_{k = 1}^K\tanh\biggl(\frac{\hbar\omega - 2\varepsilon_F^{(k)}}{4k_BT}\biggr)
\biggr\}
\end{equation}
The first two terms in the right-hand side of Eq.~(7) describe
the intraband (Drude) absorption of THz radiation in all GLs,
whereas the third term is associated with the interband transitions.
When the latter is negative, i.e., when the interband emission prevails
the interband absorption,
the  quantity Re~$\sigma_{\omega}$ as a function of $\omega$ 
exhibits a minimum. 
At a strong optical pumping when the quantities $\varepsilon_F^{(k)}$
are sufficiently large,  Re~$\sigma_{\omega} < 0$ 
in this minimum as well as
in a certain range of frequencies $\omega_{min} < \omega < \omega_{max}$. 
Here $\omega_{min}$
and $\omega_{max}$ are the frequencies at which
 Re~$\sigma_{\omega} = 0$; they are 
determined by $\tau_B$,  $\tau$, and $\varepsilon_F^{T}$
(i.e., by the intensity of the incident optical pumping radiation).

Since  in the MGL structure in question $\varepsilon_F^B \gg k_BT$, 
considering such  frequencies that 
$\omega^2\tau_B^2,\, \omega^2\tau^2 \gg 1$,
one can reduce Eq.~(7) to the following: 

$$
{\rm Re}~\sigma_{\omega} = 
\displaystyle
\biggl(\frac{e^2}{4\hbar}\biggr)\biggl\{
\displaystyle
\frac{4\varepsilon_F^{B}}
{\pi\hbar\omega^2\tau_B}
+ 
\displaystyle
\frac{8k_BT}
{\pi\hbar\omega^2\tau}\sum_{k = 1}^K
\ln \biggl[1 + \exp\biggl(\frac{\varepsilon_F^{(k)}}{k_BT}\biggr)\biggr]
$$
\begin{equation}\label{eq8}
+
 \displaystyle
\sum_{k = 1}^K\tanh\biggl(\frac{\hbar\omega - 2\varepsilon_F^{(k)}}{4k_BT}\biggr)
\biggr\}.
\end{equation}
%
%
%
%
%
%
\begin{figure}[t]
\vspace*{-0.4cm}
\begin{center}
\includegraphics[width=6.5cm]{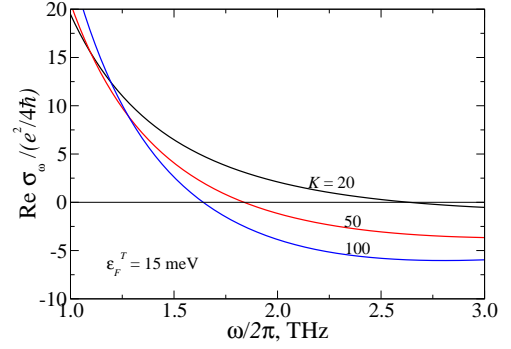}
\caption{Frequency dependences of the real part of dynamic conductivity
Re $\sigma_{\omega}$ normalized by quantity $e^2/4\hbar$
 for MGL structures with different
number of GLs $K$ at modest  pumping ($\varepsilon_F^{T} = 15$~meV.)
}
\end{center}
\end{figure} 
\begin{figure}[t]
\vspace*{-0.4cm}
\begin{center}
\includegraphics[width=6.5cm]{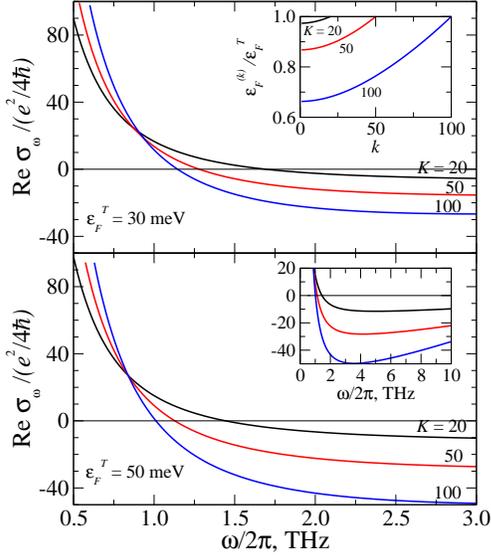}
\caption{Frequency dependences of the real part of dynamic conductivity
Re $\sigma_{\omega}$ normalized by quantity $e^2/4\hbar$
 for MGL structures with different
number of GLs $K$ at $\varepsilon_F^{T} = 30$ and 50~meV.
The inset on upper panel shows  how the  GL population varies with the 
GL index $k$, whereas the inset on lower panel
demonstrates the  dependences but  in a wider range
of frequencies.
}
\end{center}
\end{figure} 
Under sufficiently strong optical pumping when $\varepsilon_F^{(k)}  
\gg \hbar\omega, k_BT$
and, consequently, 
$\tanh[(\hbar\omega - 2\varepsilon_F^{(k)})/4k_BT] \simeq - 1$,  
setting  $\sum_{k = 1}^K \varepsilon_F^{(k)}
\simeq K^*\varepsilon_F^T$, where $K^* <  K$,
from Eq.~(8)
one obtains
\begin{equation}\label{eq9}
{\rm Re} \sigma_{\omega} \biggl(\frac{4\hbar}{e^2}\biggr)\simeq 
\displaystyle
\frac{4}{\pi\hbar\omega^2}
\biggl(\frac{\varepsilon_F^B}{\tau_B} + 
\frac{2K^*\varepsilon_F^T}{\tau}\biggr) - K .
\end{equation}

\begin{figure}[t]
\vspace*{-0.4cm}
\begin{center}
\includegraphics[width=6.5cm]{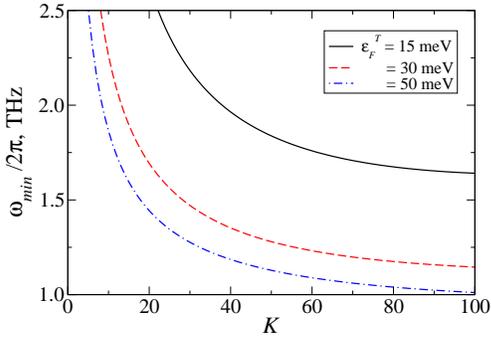}
\caption{ Dependences of $\omega_{min}$  on 
number of GLs $K$ at  different $\varepsilon_F^T$.
}
\end{center}
\end{figure} 

\begin{figure}[t]
\vspace*{-0.4cm}
\begin{center}
\includegraphics[width=6.5cm]{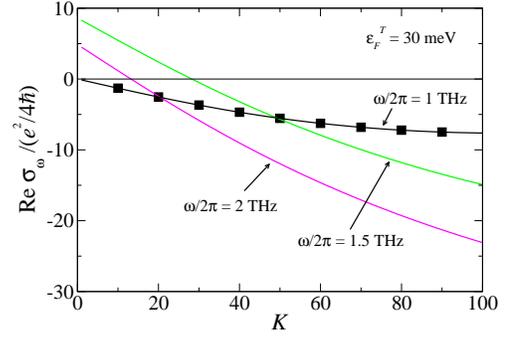}
\caption{Real part of the normalized dynamic conductivity as a function of
the number of GLs for  different frequencies.
The line with markers corresponds to a MGL structure without the bottom GL.
}
\end{center}
\end{figure}

\section{Frequency characteristics of dynamic conductivity}

Figures~2 and 3 show the frequency dependences of ${\rm Re}~\sigma_{\omega}$
normalized by $e^2/4\hbar$
calculated for MGL structures with different
$K$ at different values of $\varepsilon_F^T$ (i.e., different optical pumping 
intensities)
using Eqs.~(1), (2), (5), and (6) or Eqs.~(5) and (7). 
We set $\varepsilon_F^B = 400$~meV~\cite{8},
$\hbar\Omega = 920$~meV, $T = 300$~K, $\tau_B = 1$~ps, $\tau = 10$~ps, and $\gamma = 1/4$.
As seen from  Figs.~2 and 3,  ${\rm Re}~\sigma_{\omega}$ can be negative
in the frequency range $\omega > \omega_{min}$ with $\omega_{min}$ decreasing
with increasing quasi-Fermi energy $\varepsilon_F^T$ in the topmost GL,
i.e., with increasing optical pumping intensity (see below). 
In the MGL structures with $K = 50 - 100 $
at  $\varepsilon_F^T = 30 - 50$~meV, one has
$\omega_{min}/2\pi \gtrsim 1$~THz (see Figs.~3 and 4). 
As follows from the inset on upper panel in Fig.~3, the quantities
$\varepsilon_F^{(k)}$ are not too small in comparison with $\varepsilon_F^T$
even in GLs with the  indices $k \ll K$, i.e., in GLs near the MGL 
structure bottom.
This implies that the pumping of such near bottom GLs is effective even 
in the MGL structures with $K \sim 100$.
The quantity  ${\rm Re}~\sigma_{\omega}$ as a function of $\omega$ exhibits a minimum
(see the inset on lower panel in Fig.~3).
The sign of  ${\rm Re}~\sigma_{\omega}$ becomes positive at 
$\omega > \omega_{max}$,
where $\omega_{max}$ is  rather large: more than 10~THz.
The MGL structures with larger number $K$ of GLs  at stronger pumping
exhibit smaller $\omega_{min}$
and deeper minima ${\rm Re}~\sigma_{\omega}$.
This is confirmed also by Fig.~5.

\section{Role of the bottom GL}

The highly conducting bottom GL, whose existence is associated with
the intrinsic features of the MGL structure growth, plays the negative role.
This is mainly because it results in a marked absorption (due to the Drude
mechanism) of the THz radiation
emitted by other GLs. Since such an absorption increases with 
decreasing frequency,
the achievement of the negative dynamic conductivity in the MGL structures with
shorter momentum relaxation times $\tau_B$ and $\tau$ even at higher
frequencies is significantly complicated by the Drude absorption.
This is demonstrated by
Fig.~6 (upper panel).
In contrast, 
a decrease in the Fermi energy $\varepsilon_F^B$ and  the electron density
in the bottom GL, of course, might significantly promote the achievement
of negative dynamic conductivity in a wide frequency range, particularly,
at relatively low frequencies (compare the frequency 
dependences on upper and lower panels in 
Fig.~6).
In this regard,  the MGL structures with  GLs exhibiting a long relaxation
time $\tau$
(like that found  in Ref.~\cite{10}) but with  a lowered electron density 
in the bottom
GL or  without the bottom layer 
appears to be much more preferable. 
Such MGL structures can be fabricated using 
chemical/mechanical reactions and transferred substrate techniques
(chemically etching the substrate and the highly conducting bottom 
GL~\cite{12} or mechanically 
peeling the  upper GLs, then transferring the upper portion of
the  MGL structure on a Si or equivalent transparent substrate).
The calculation of Re~$\sigma_{\omega}$ for this MGL structure can be carried out
by omitting the term Re~$\sigma_{\omega}$ in Eq.~(6).
The pertinent results are shown in Fig.~5 (see the marked line) and 
Fig.~7. 
Here as in Fig.~3, we assumed that
that
$\hbar\omega = 920$~meV, $T = 300$~K, $\tau_B = 1$~ps, and $\tau = 10$~ps.
\begin{figure}[t]
\vspace*{-0.4cm}
\begin{center}
\includegraphics[width=6.5cm]{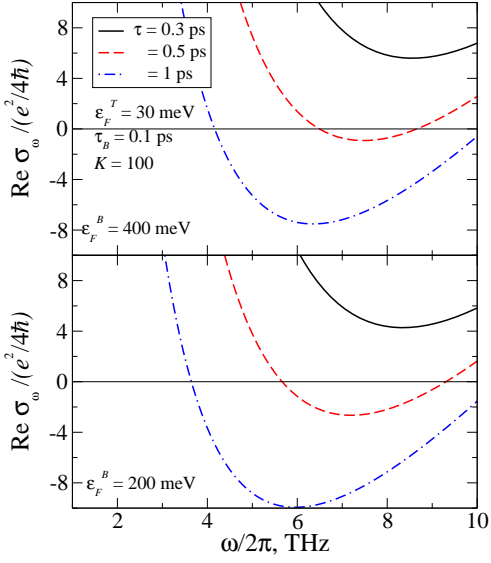}
\caption{Real part of the normalized dynamic conductivity versus frequency
calculated for MGL structures with $\tau_B = 0.1$~ps
and different $\tau$ and $\varepsilon_F^B$.
}
\end{center}
\end{figure}
\begin{figure}[t]
\vspace*{-0.4cm}
\begin{center}
\includegraphics[width=6.5cm]{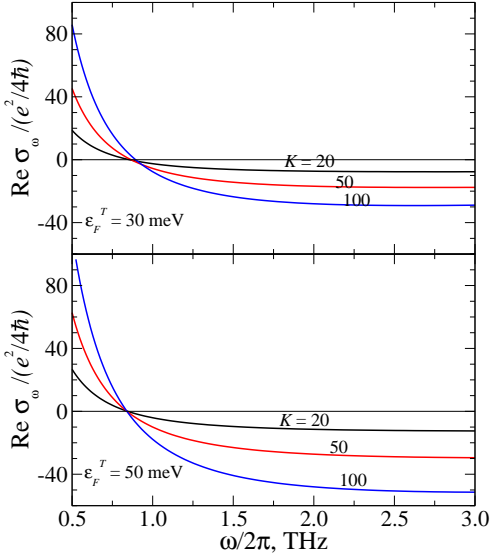}
\caption{Real part of the normalized dynamic conductivity versus frequency
calculated for MGL structures with different numbers of GLs and without bottom GL.
}
\end{center}
\end{figure}

The obtained frequency
dependences are qualitatively similar to those shown in Figs.~2 and 3. However the 
dependences for the MGL structures without the bottom layer exhibit a marked shift toward lower frequencies. In particular, as follows from Figs.~5
and 7,  Re~$\sigma_{\omega}$ can be negative even at $\omega/2\pi \lesssim 1$~
THz (at choosen values
of $\varepsilon_F^T$).
 
\section{Condition of lasing}

To achieve  lasing in the MGL structures under consideration,
the following condition should be satisfied~\cite{7}:
\begin{equation}\label{eq10}
\frac{8\pi}{c}|{\rm Re}~\sigma_{\omega}|\,E^2_t > 
(1 - r_1)E_1^2 + (1 - r_2)E_2^2 + (a/R)^2E_1^2 + E_S^2.
\end{equation}
Here, $E_t$, $E_1$, and $E_2$ are maximum amplitudes of the THz electric field
$E = E(z)$ at the 
MGL structure (placed at the distance $t$ from the the bottom mirror, 
where $t$ 
is the thickness of the substrate),
and near the pertinent mirror, respectively, 
$E_S^2 = (\alpha_Sn_S/2)\int_0^{t}E^2dz$, $\alpha_S$ and $n_S$ are the 
absorption coefficient of THz radiation in the substrate (SiC or Si) and real part of its
refraction index, $r_1$ and $r_2$ are the reflection coefficients of THz radiation
from the mirrors, and $a/R$ is the ration of the diameters of 
the output hole $a$
and the mirror $R$. 
In deriving inequality~(10), we neglected the finiteness 
of the MGL thickness (in comparison with $t$ and the THz wavelength) and 
disregarded
the diffraction losses.
For simplicity, one can set $E_t^ \sim E_1^2 \sim E_2^2$ and $E_S^2 \sim 
(t\alpha_Sn_S/2)E_2^2$, disregarding, in particular, partial
 reflection of THz radiation from the bottom GL.
In this case, inequality~(10)
can be presented as
\begin{equation}\label{eq11}
\frac{8\pi}{c}|{\rm Re}~\sigma_{\omega} | > (1 - r_1) + (1 - r_2) + (a/R)^2 + 
t\,\alpha_Sn_S/2 = L.
\end{equation}
Assuming that $r_1 = r_2 = 0.99$, $(a/R) = 0.1$, 
$\alpha_S \simeq 2 - 4$~cm$^{-1}$, $n_S \simeq 3$~\cite{13},
and $t = 50~\mu$m, 
for $L$ one obtains $L = 0.06 - 0.09$. 
However, as follows from Fig.~5, 
the quantity $(8\pi/c)|{\rm Re}~\sigma_{\omega}|$ for a MGL structure with
$K = 100$ at $\omega/2\pi = 1.5$~THz at $\varepsilon_F^T = 30$~meV
is about $12\beta \simeq 0.275$.
At $\omega/2\pi = 1.0$~THz but for the structure without the bottom GL
(see the line with markers in Fig.~5), one obtains  
$(8\pi/c)|{\rm Re}~\sigma_{\omega}| \simeq 0.345$. These values
of  $(8\pi/c)|{\rm Re}~\sigma_{\omega}|$ well exceed 
the above value of $L$ (in contrast with the structures with two GLs~\cite{7},
for which the minimization of the losses is crucial).
At the elevated frequencies, 
the ratio $(8\pi/c)|{\rm Re}~\sigma_{\omega}|/L$
can be even much larger.

Considering that the electron (hole) density
$\Sigma^T$ in the topmost GL 
$$
\Sigma^T = \frac{2}{\pi\hbar^2}\int_0^{\infty}
\frac{dp\,p}{1 + \exp[(v_Fp - \varepsilon_F^T)/k_BT]}
$$
at  $\varepsilon_F^T = 30$~meV at $T = 300$~K,
we obtain
$\Sigma^T \simeq 2\times 10^{11}$~cm$^{-2}$.
Such a value (and higher) of the photogenerated electron and hole density
 is achievable experimentally (see, for instance,
Ref.~\cite{14}).
Assuming that $K = 100$, $\hbar\Omega = 920$~meV, and the recombination time
$\tau_R \simeq 20$~ps at $T = 300$~K 
and $\Sigma^T \simeq 2\times 10^{11}$~cm$^{-2}$~\cite{15}, 
for the pertinent optical pumping power
we obtain $I_{\Omega} \simeq  6.4\times 10^4 $~W/cm$^2$.
One needs to point out that this value of  $I_{\Omega}$ is much larger
than the threshold of lasing at a certain frequency ($\omega_{min} < \omega < \omega_{\max}$).
At $T = 100$~K, the recombination time (due to optical phonon emission)
is much longer~\cite{15}, hence,
the electron and hole densities in question can be achieved at much 
 weaker optical pumping. 

In the regime sufficiently beyond  the threshold of lasing when
the stimulated radiative recombination becomes  dominant,
the pumping efficiency is determined just by the ratio of
the energy of the emitted THz photons $\hbar\omega$ and the energy of optical photons 
$\hbar\Omega$: $\eta = \omega/\Omega$.
However, the nonradiative recombination mechanisms can markedly 
decrease $\eta$.
In the regime in question, the maximum
output THz power can be estimated as
${\rm max}P_{\omega} \simeq \pi\,R^2(\omega/\Omega)I_{\Omega}$.
For example, at $\hbar\Omega = 920$~meV,
$\hbar \omega/2\pi = 5.9$~meV$ (\omega/2\pi \simeq 1.5$~THz), $2R = 0.1$~cm,
and  
 $I_{\Omega} =  3\times10^4$~W/cm$^2$, one obtains
max~$P_{\omega} \simeq 1.5$~W.


\section{Conclusions}

We studied  real part of the dynamic conductivity of 
a  multiple-graphene-layer (MGL) structure with a stack of GLs and a highly 
conducting
bottom GL 
on SiC substrate pumped by optical radiation. It was shown that
 the negative dynamic conductivity in the MGL structures 
under consideration with sufficiently large number $K$ of
perfect upper GLs can be achieved
even at room temperature provided  the optical pumping is sufficiently strong.
Due to large $K$, the absolute value of Re$\sigma_{\omega}$ in its minimum
can significantly exceed the characteristic value of conductivity $e^2/4\hbar$.
Thus,  the MGL structures can serve active media of THz lasers.
This can markedly liberate the requirement for the quality of
the THz laser  resonant cavity.
An increase in $\tau_B$ and $\tau$ promotes widening of the frequency range
where Re$\sigma_{\omega} < 0$, particularly, at the low end of this range.
This opens up the prospects of 
THz lasing with  $\omega/2\pi \sim 1$~THz even at room temperature.
The main obstacle appears to  be  the necessity of sufficiently long
relaxation times $\tau_B$ and $\tau$ in GLs with rather high electron
and hole densities: $\Sigma >  10^{11}$~cm$^{-2}$.
Since the electron-hole recombination at the temperatures and densities
under consideration
might be attributed to the optical phonon emission~\cite{16}, 
the  optical pumping intensity required for THz lasing 
can be markedly lowered (by orders of magnitudes)
 with  decreasing temperature.
The conditions of lasing in the MGL structures  without the bottom GL
are particularly liberal.

\section*{Acknowledgments}

One of the authors (V.R) is grateful to M.~Orlita and F.~T.~Vasko
for useful discussions and 
information.
This work was supported by the Japan Science and technology Agency, CREST, 
Japan.
Partially it was also supported by the Russian Academy of Sciences, Russia.

\section*{Appendix A. Recombination}

\setcounter{equation}{0}
\renewcommand{\theequation} {A\arabic{equation}}

The rate of radiative recombination  in the degenerate electron-hole system in the
topmost 
GL due to spontanious emission
of photons at $\varepsilon_F^T \gg k_BT$
can be calculated using the following formula~\cite{6,15}
\begin{equation}\label{A1}
R_r \simeq \frac{2v_r}{\pi\hbar^3}\int_0^{\varepsilon_F^T/v_F}dp\,p^2 =
\frac{2v_r(\varepsilon_F^T)^3}{3\pi\hbar^3v_F^3} \propto (\varepsilon_F^T)^3.
\end{equation}
Here $v_r = \sqrt{\ae}(8e^2/3\hbar\,c)(v_F/c)^2v_F$ and 
$\ae$ is the dielectric constant $p_F = \varepsilon_F^T/v_F$.

The rate of the electron-hole recombination associated with emission of optical phonons
can be  described the following equation~\cite{14}:
$$
R_{ph} \propto \int_0^{\hbar\omega_0/v_F}dp\,p (\hbar\omega_0/v_F - p)
$$
$$
\times \biggl[1 + \exp\biggl(\frac{v_Fp - \varepsilon_F^T}{k_BT}\biggr)\biggr]^{-1}
\biggl[1 + \exp\biggl(\frac{\hbar\omega_0 - v_Fp - \varepsilon_F^T}{k_BT}\biggr)\biggr]^{-1}
$$
\begin{equation}\label{A2}
\propto \exp\biggl(\frac{2\varepsilon_F^T - \hbar\omega_0}{k_BT}\biggr).
\end{equation}

Equalizing $R_r^T$ and the rate of generation of electrons and holes
by the optical pumping radiation $G_{\Omega}^T$ and the recombination rate,
and considering Eqs.~(A1) and (A2), 
one can find that $\varepsilon_F^T \propto (G_{\Omega}^T)^{1/3}$ and
 $\varepsilon_F^T \propto \ln(G_{\Omega}^T)$, respectively.
The radiative interband transitions stimulated by the thermal photons can also contribute
to the recombination rate~\cite{6,15} as well as the processes of electron-hole interaction 
in the presence of disorder. These processes provide different dependences of the recombination
rate on the quasi-Fermi energy. Due to this, calculating the dependence of the latter
in different GLs on the optical pumping intensity, 
we put, for definiteness,  $\varepsilon_F^T 
\propto (G_{\Omega}^T)^{\gamma}$ with $\gamma = 1/4$. Such a dependence is less steep
than that for the spontaneous radiative recombination and somewhat steeper
than in the case of optical phonon recombination. 
The variation of parameter $\gamma$, leads to some change in the $\varepsilon_F^{(k)} - k$ 
dependence, but it should not affect the main obtained results.


\begin{references}
\bibitem[*]{byline}  Electronic address: v-ryzhii@u-aizu.ac.jp


\bibitem{1}
C.~Berger, Z.~Song, T.~Li, X.~Li, A.Y.~Ogbazhi, R. Feng,
Z.~Dai, A.~N.~Marchenkov, E.~H.~Conrad, P.~N.~First, and W.~A.~de Heer,
J.~Phys. Chem. {\bf 108}, 19912 (2004). 
%

\bibitem{2}
K.~S.~Novoselov, A.~K.~Geim, S.~V.~Morozov, D.~Jiang,
M.~I.~Katsnelson, I.~V.~Grigorieva, S.~V.~Dubonos, and A.~A.~Firsov,
Nature {\bf 438}, 197 (2005).
%


\bibitem{3}
V.~Ryzhii, M.~Ryzhii, and T.~Otsuji,
J. Appl. Phys.  {\bf 101} (2007) 083114.
%
\bibitem{4}
M.~Ryzhii and V.~Ryzhii,
Jpn. J. Appl. Phys. {\bf46 }, L151 (2007).

\bibitem{5}
F.~Rana, IEEE Trans. Nanotechnol. {\bf 7}, 91 (2008)
 \bibitem{6}
A.~Satou, F.~T.~Vasko, and V.~Ryzhii, 
Phys. Rev. B {\bf 78},  115431 (2008). 

\bibitem{7} 
A.~A.~Dubinov, V.~Ya.~Aleshkin,
M.~Ryzhii, T.~Otsuji,
and V.~Ryzhii,
Appl. Phys. Expess (in press). 

%
\bibitem{8}
F.~Varchon, R.~Feng, J.~Hass, X.~Li, B.~Ngoc Nguyen, C.~Naud, P.~Mallet, J.-Y.~Veuillen, C.~Berger, E.~H.~Conrad, and L.~Magaud,
Phys. Rev. Lett. {\bf 99}, 126805 (2007).
\bibitem{9}
M.~Orlita, C.~Faugeras, P.~Plochocka, P.~Neugebauer, G.~Martinez, D.~K.~Maude, A.-L.~Barra, M.~Sprinkle,
C.~Berger, W.~A.~de Heer, and M.~Potemski,
Phys. Rev. Lett. {\bf 101}, 267601 (2008).

\bibitem{10}
P.~Neugebauer, M.~Orlita, C.~Faugeras,  A.-L.~Barra, and M.~Potemski,
Anstracts of 14th Int. Conf. Narrow Gap Semicon. and Systems, p.~34.
Sendai, Japan, July 13-17 (2009); arXiv: 0903.1612 [cond-mat.mes-hall] (2009).


%
\bibitem{11}
L.~A.~Falkovsky and A.~A.~Varlamov, Eur. Phys. J. B {\bf 56}, 281 (2007)
 
\bibitem{12}
A.~Bostwick, T.~Ohta, T.~Seyller, K.~Horn, and E.~Rotenberg,
Nature Phys. {\bf 3}, 36 (2007). 

\bibitem{13}
J.~H.~Strait,P.~A.~George, J.~M.~Dawlaty, S. Shivaraman, M. Chanrashekhar,
F.~Rana, and M.~G.~Spencer,
Appl. Phys. Lett. {\bf 95}, 051912 (2009).
\bibitem{14}
J.~M.~Dawlaty, S.~Shivaraman, M.~Chandrashekhar, F.~Rana, and M.~G.~Spencer,
Appl. Phys. Lett. {\bf 92}, 042116 (2008). 



\bibitem{15}
F.~Rana, P.~A.~George, J.~H.~Strait, S. Shivaraman,
M. Chanrashekhar, and M.~G.~Spencer,
Phys. Rev. B {\bf 79}, 115447 (2009).
\bibitem{16}
F.~T.~Vasko and V.~Ryzhii,
Phys. Rev. B {\bf 77}, 195433 (2008).

\end{references}
\end{document}